\documentstyle{elsart}

\begin{document}
\begin{frontmatter}

\title{Experimental evidence of multiple magnetic relaxation processes in
Mn$_{12}$ acetate and Mn$_{12}$~2-Cl benzoate}

\author{Marco Evangelisti}, 
\author{Juan Bartolom\'e\thanksref{cor}} and 
\author{Fernando Luis\thanksref{lei}}
\address{Instituto de Ciencia de Materiales de Arag\'on,   
C.S.I.C. - Universidad de Zaragoza, E-50009 Zaragoza, Spain}
\thanks[cor] {Corresponding author. Tel. +34 976 761218; Fax +34 976 761229;
E-mail barto@posta.unizar.es}
\thanks[lei] {On leave of absence at the Kamerlingh Onnes Laboratory, Leiden
University, PO Box 9506, NL-2300 RA Leiden, The Netherlands}

\begin{abstract}
The Mn cluster complexes Mn$_{12}$(RCOO)$_{16}$(H$_{2}$O)$_{4}$O$_{12}$ with
R~=~CH$_{3}$ and 2ClPh (hereafter referred to as Mn$_{12}$ acetate and
Mn$_{12}$~2-Cl benzoate, respectively) exhibit macroscopic magnetic quantum
tunneling. Dynamical magnetic measurements indicate for both samples that the
anisotropy energy barrier is nearly the same and that the relaxation times
have sharp minima at the same resonant fields. However, we show that for both
compounds a minority portion of the clusters does not undergo this main
relaxation process. From ac-susceptibility experiments, we report evidence of
a continuous distribution of faster relaxation times, giving rise to a second
relaxation regime. The zero-field effective relaxation times of such a
distribution follow an Arrhenius law with energy barrier of approximately 23
and 30~K for the acetate and the benzoate, respectively. In the case of
Mn$_{12}$~2-Cl  benzoate, the field dependence of the effective relaxation
time points towards the existence of spin tunneling with $H_{z}=0$ being a
resonant field. The existence of a broad distribution of energy barriers is
corroborated in magnetization experiments by the observation of logarithmic
relaxation at short times.
\end{abstract}
\end{frontmatter}

At present there is a considerable interest in Mn$_{12}$ complexes because
they can exhibit macroscopic quantum tunneling of the magnetic moment (QTMM).
The most thoroughly studied Mn$_{12}$ complex exhibiting QTMM is the commonly
known as Mn$_{12}$ acetate, first synthesized by Lis~\cite{lis}. The
large-spin ($S=10$) ground state together with an appreciable magnetic
anisotropy results in a barrier for reversal of the direction of
magnetization. Steps in the hysteresis loop~\cite{mag} and sharp minima in the
field dependence of the relaxation time deduced from ac-susceptibility
measurements~\cite{acx} have proven the existence of magnetic quantum
tunneling processes in Mn$_{12}$ acetate. However, low temperature
magnetization experiments in Mn$_{12}$ acetate have given contradictory
results concerning mainly the relaxation rate~\cite{testo,paulsen,thom,pere}.
Several authors have pointed out that there is a magnetic minority component
in almost all Mn$_{12}$ complexes, different from the well studied main
component, which undergoes fast relaxation in the 2~K
region~\cite{chem1,chem2,fluis,chem4}. Since any relaxation observation is
dominated at each temperature by the fastest process, it is of paramount
importance to ascertain which is the magnetic species that is observed in a
particular relaxation experiment. Our goal is to clarify this point in the
temperature region of 2~K, where the minority species relaxes most
effectively, and which is the dominant mechanism of the relaxation process.

In this letter, we present low temperature ac and dc experiments on oriented
crystallites of [Mn$_{12}$(CH$_{3}$COO)$_{16}$\-(H$_{2}$O)$_{4}$O$_{12}$]$\-
\cdot$CH$_{3}$COOH$\cdot$4H$_{2}$O, that is the commonly known as Mn$_{12}$
acetate, and on single-crystal
[Mn$_{12}$\-(2ClPh\-COO)$_{16}$\-(H$_{2}$O)$_{4}$O$_{12}$]$\-
\cdot$CH$_{2}$Cl$_{2}\cdot$5H$_{2}$O, hereafter referred to as Mn$_{12}$~2-Cl
benzoate~\cite{chem3}. The magnetic core and the local symmetry of each
molecule are the same in both compounds. However, in contrast to Mn$_{12}$
acetate, the easy axes of Mn$_{12}$~2-Cl benzoate molecules lie alternatively
on the directions [011] and [0$\bar{1}$1], being nearly perpendicular to their
nearest neighbors. The magnetic measurements here presented were performed in
a commercial SQUID magnetometer with an ac-susceptibility measurement option.
The external field and the ac-field were applied along the easy axes of the
molecules in the Mn$_{12}$ acetate and along one of the [011] and
[0$\bar{1}$1] directions, that is, along the easy axes of one half of the
molecules in the Mn$_{12}$~2-Cl benzoate. The amplitude of the ac-field was 
4.5~Oe in all the experiments and the frequency $f$ was varied between 0.1 and
990~Hz. In the relaxation experiments, the samples were firstly saturated
with a magnetic field of 5~T. Then the field was reduced to zero and the
magnetization started to be recorded with a delay time of about 5 minutes.

Figure~1 shows the out-of-phase ac-susceptibility of Mn$_{12}$ acetate and
Mn$_{12}$~2-Cl benzoate as a function of the temperature. For both samples, a
clear peak is observed in the temperature range 4 to 9~K. A detailed
experimental study of the magnetic relaxation in this temperature range has
been presented by some of us in Ref.~\cite{acx} for the acetate and in
Ref.~\cite{mncl2} for the benzoate. One of the main findings is that, at zero
applied field, the temperature dependence of the relaxation time follows an
Arrhenius law, $\tau=\tau_{0}~{\rm{exp}}~(U/k_{\rm{B}}T)$, with
$U\approx 65$~K for both compounds and $\tau_{0}\approx 3\times 10^{-8}$~s for
the acetate and $\tau_{0}\approx 10^{-8}$~s for the benzoate. At fields below
1~KOe, it is also shown that in the benzoate the tunneling takes place
approximately through the same states as in the acetate.

As shown in the inset of Fig.~1, another peak can be observed at temperatures
below 4~K for both samples. Even though its height is an order of magnitude
smaller than the height of the main peak, it is well defined. The existence of
two peaks in the out-of-phase ac-susceptibility of Mn$_{12}$ complexes has
been already observed~\cite{chem1,chem2,fluis,chem4}. Notably, the relative
heights depend strongly on the sample batch and history. However, to the best
of our knowledge, a detailed study of the frequency dependence of the smaller
peak is missing in the literature. In the rest of this letter, we will focus
on the magnetic relaxation which gives rise to this smaller peak.

If the thermal energy is less than the effective energy barrier for reversal
of magnetization direction, then a maximum in the out-of-phase
ac-susceptibility is seen when the frequency equals the rate of magnetization
reversal. The relaxation time constant $\tau$ can be determined from the
measured susceptibilities by plotting $\chi\prime\prime(\omega)$ versus
$\chi\prime(\omega)$ in so-called Argand diagrams~\cite{argand}. This plot
yields a semi-circle if only one relaxation time is involved. The relaxation
time is found as the inverse angular frequency at the top of the semi-circle.
Figure~2 shows an example for each sample of zero-field Argand diagram of data
collected at  $T=1.7$~K, that corresponds to the region of the smaller peak of
Fig.~1. The continuous lines are the fits to a semi-circle dependence. In the
case of the Mn$_{12}$~2-Cl benzoate, at high frequencies, the experimental
data deviate from the fit and increase, indicating the presence of, at least,
another faster relaxation process at frequencies exceeding the limit of our
instrument. On the contrary in the case of the Mn$_{12}$ acetate, the
frequency window of our instrument allows us to fit to a semi-circle the data
corresponding to higher frequencies, while a strong deviation is found at low
frequencies. Moreover, in both cases, most of the data lie in a strongly
flattened semi-circle. Such a behaviour is typical of a continuous
distribution of semi-circles and, consequently, of a continuous distribution
of relaxation times. Effective time constants can be obtained from the
reciprocal angular frequency at maximum absorption, but one has to realize
that these times give only an average value~\cite{argand}. The zero-field
relaxation times so obtained together with the ones of Ref.~\cite{mncl2} are
collected in Fig.~3. In the temperature range 1.7 to 2.2~K, the relaxation
times can be approximated by an Arrhenius law with characteristic energy
$U\approx23$~K and $\approx30$~K and pre-exponential factor $\tau_{0}\approx
10^{-9}$~s and $\approx 2\times 10^{-9}$~s for the acetate and the benzoate,
respectively.

The field dependence of the effective relaxation times is particularly
interesting. The experimental points of Fig.~4 have been obtained, as
previously done, by fitting the Argand diagrams of the data measured at
different fields and at $T=1.7$~K (acetate) or $T=2.0$~K (benzoate). Again,
evidences of slower (for the acetate) and faster (for the benzoate) relaxation
processes are found for all applied fields. The striking feature of Fig.~4 is
that the relaxation times of Mn$_{12}$~2-Cl benzoate are not simply decreasing
as expected from a classical behaviour. In contrast, they increase for fields
up to 750~Oe and decrease for higher fields. The imaginary component of the
ac-susceptibility found at fields higher than 1.5~KOe is too small to obtain
reliable results from the analysis of Argand diagrams.

The experimental data taken from the frequency window of our
ac-mea\-su\-re\-ments, thus far reported, can be summarized in the following
way. First, for $3.8~{\rm{K}}<T<6.0~{\rm{K}}$, a large portion of the
total magnetization relaxes exponentially with a characteristic time 
$\tau_{1}\propto{\rm{exp}}~(U_{1}/k_{\rm{B}}T)$ varying between 1~s and
$10^{-3}$~s, respectively, for the acetate, and $4\times 10^{-1}$~s and
$5\times 10^{-4}$~s, respectively, for the benzoate, and with an effective
energy barrier $U_{1}\approx 65$~K for both samples. Second, for
$1.7~{\rm{K}}<T<2.0~{\rm{K}}$, a small portion of the total magnetization
relaxes exponentially with an effective relaxation time 
$\tau_{2}\propto{\rm{exp}}~(U_{2}/k_{\rm{B}}T)$ varying between
$2\times 10^{-3}$~s and $6\times 10^{-5}$~s, respectively, for the acetate,
and $10^{-1}$~s and $5\times 10^{-3}$~s, respectively, for the benzoate, and
with effective energy barriers $U_{2}\approx 23$~K for the acetate and
$U_{2}\approx 30$~K for the benzoate. Third, by applying a field,
$\tau_{2}(H_{z})$ in the benzoate has a maximum at $H_{z}\neq0$.

In the case of Mn$_{12}$~2-Cl benzoate, the behaviour of $\tau_{2}(H)$ at
$T=2.0$~K (Fig.~4) provides clear evidence of the existence of quantum
tunneling between excited states with $H_{z}=0$ being a resonant field. When
the field increases, the tunneling becomes progressively blocked and,
consequently, the relaxation becomes slower. The field dependence of the
energy barrier for uniaxial magnetic anisotropy is given by 
$U(H)/U(0)=(1-g\mu_{\rm{B}}H/2DS)^{2}$, where $D$ is the anisotropy energy
constant. For a comparison, Fig.~4 shows also the estimated dependence of the
relaxation time $\tau_{2}(H)=\tau_{0}~{\rm{exp}}~(U_{2}(H)/k_{\rm{B}}T)$ for
classical over-barrier hopping calculated for $\tau_{0}=2\times 10^{-9}$~s,
$D=0.35$~K and $S=10$, that gives rise to a zero-field energy barrier
$U_{2}(0)=DS^{2}=35$~K. The fact that the experimental data deviate from the
theoretical curve at higher fields points towards the existence of another
minimum and, consequently, to another resonant field. On the contrary in the
case of the Mn$_{12}$ acetate, no evidence of magnetic quantum tunneling is
found from the field dependence of $\tau_{2}$ (Fig.~4). The data are well
described by classical over-barrier hopping with $\tau_{0}=10^{-9}$~s,
$D=0.25$~K and $S=10$.

As outlined above, the presence of secondary fast relaxation processes may be
associated to a minority portion of the Mn$_{12}$ clusters. A rough estimate
of the relative number of clusters experiencing faster relaxation can be
deduced by considering the out-of-phase susceptibility given by
\begin{equation} 
\chi\prime\prime(\omega,T)=
\frac{n_{1}~(\chi_{\mathrm{T}}^{(1)}-\chi_{\mathrm{S}}^{(1)})~\omega\tau_{1}}
{1+\omega^{2}\tau_{1}^{2}}~+~
\frac{n_{2}~(\chi_{\mathrm{T}}^{(2)}-\chi_{\mathrm{S}}^{(2)})~\omega\tau_{2}}
{1+\omega^{2}\tau_{2}^{2}},
\end{equation}
where the indices 1 and 2 refer to the majority and minority species,
respectively, and $n_{i}$ is the relative amount of the species $i$,
$\omega=2\pi f$, $n_{2}=1-n_{1}$, 
$\tau_{i}=\tau_{0}^{(i)}{\rm{exp}}(U_{i}/k_{\mathrm{B}}T)$ and
$\chi_{\mathrm{T}}^{(i)}$ and $\chi_{\mathrm{S}}^{(i)}$ are the isothermal and
the adiabatic susceptibility, respectively. Plotting Eq.~(1) as a function of
the temperature, $\chi\prime\prime(\omega,T)$ has two peaks centered at
$T_{\mathrm{max}}^{(i)}$, where 
$\omega\tau_{i}(T_{\mathrm{max}}^{(i)})\approx 1$, with $i=1,2$.
Considering the experimental values of
$\chi\prime\prime(\omega=2\pi\cdot620,T_{\mathrm{max}}^{(i)})$ and the
estimated $\tau_{i}(T_{\mathrm{max}}^{(i)})$, and assuming, as a first
approximation, that  
$(\chi_{\mathrm{T}}^{(1)}-\chi_{\mathrm{S}}^{(1)})\cdot T_{\mathrm{max}}^{(1)}
=(\chi_{\mathrm{T}}^{(2)}-\chi_{\mathrm{S}}^{(2)})\cdot
T_{\mathrm{max}}^{(2)}$, we obtain, using Eq.~(1), $n_{2}$(acetate)$\approx
0.05$ and $n_{2}$(benzoate)$\approx 0.01$, that is approximately 5\% and 1\%
of all clusters of acetate and benzoate, respectively, do not undergo the main
relaxation process. 

The presence of such fast processes may pass undetected in magnetization
measurements, since, in a typical experiment, the relaxation of the
magnetization begins to be recorded only after a time that is longer than the
decay times of the secondary relaxation processes. However, careful relaxation
measurements close to the saturation show that, during approximately the first
1000~s at 2.0~K and up to 2000~s at 1.7~K, the data are well fitted by
straight lines if plotted as a function of the logarithm of the
time~(Fig.~5). Logarithmic decay of the magnetic moment can be explained in
terms of a continuous distribution of energy barriers~\cite{teja}. In this
way, we have again evidence of a minority species relaxing faster and with
different $DS^{2}$ as observed in the frequency window of our ac-experiments.
Indeed, in Mn$_{12}$ acetate at temperatures lower than 2~K, evidence of
faster relaxations has been detected in Ref.~\cite{thom} by the extrapolated
zero-time magnetizations which turn out to be temperature dependent. While, at
longer times the relaxation follows a square root time dependence as we have
observed in both samples~(insets of Fig.~5).

Recently, very low temperature magnetization experiments have shown that the
relaxation of Mn$_{12}$ acetate has logarithmic time dependence.~\cite{pere}.
Such a behaviour was explained considering that the dipolar field is changing
as the system relaxes. It was argued that the dipolar field of tunneled spins
remove from resonance excited states close to the top of the energy barrier.
On the other hand, we underline that the main effect of dipolar fields is
rather to hinder the tunneling through the lower energy states that have a
tunneling splitting much smaller than that of the excited states closer to
the top of the barrier. Therefore, the mechanism of dipolar fields would be
most effective in destroying tunneling near resonance where several tunneling
channels contribute to the relaxation. Even though the relaxation proceeds
through several resonant channels, only one relaxation time can be associated
to each molecule~\cite{julio}. A distribution of relaxation  times could then
be associated to different clusters experiencing slightly different local
fields. Let us suppose that the secondary peak, related to the faster
relaxation process, observed in the out-of-phase susceptibility, arises from
those clusters in which the small local field allows the tunelling to proceed
through lower energy states. Then, this secondary peak should be much more
sensitive to an applied field than in the case when the peak arises from the
tunneling of a minority species with different $DS^{2}$. In fact, by
destroying the resonance, the field should shift the secondary peak towards
the main peak at higher temperature. However, we observe that the secondary
peak in the out-of-phase susceptibility does not move to higher temperature
when a field of 500~Oe is applied, that is, when the system is off resonance.
As a consequence, it seems more probable that the origin of the logarithmic
decay can be associated to the existence of minority species having different
energy barriers.

What could be the possible origin of these secondary species? A first
possibility could be due to a minority portion of the Mn$_{12}$ clusters with
a different value of $S$ for the ground state. There has been some evidence
for $S=9$ Mn$_{12}$ complexes from variable-field magnetization
studies~\cite{chem4}. In this case, from the zero-field splitting $DS^{2}$,
the effective energy barrier decreases. However, taking $D=0.6$~K as estimated
from ESR experiments~\cite{esr} for the clusters relaxing around 5~K, we
obtain an energy barrier much higher than the one that we estimate in this
paper. Consequently, we should expect either a spin value $S<9$, or, as a
second possibility, a lower axial anisotropy value. The latter may be the case
if one of the Mn ions presents the Jahn-Teller distortion along a direction
perpendicular to the main anisotropy axis~\cite{jt}. Both possibilities can
coexist and perhaps arise from clusters lying near the surface. Therefore,
the results here presented may depend strongly on sample shape and dimension.
In this scenario, the greater relative amount of the secondary species and
its broader distribution of relaxation times observed in the sample of
acetate, constituted by oriented crystallites, than those observed in the
single-crystal sample of benzoate, can be naturally explained.  

In conclusion, we have reported direct evidence of multiple processes of
magnetic relaxation in Mn$_{12}$ acetate and Mn$_{12}$~2-Cl benzoate by
ac-susceptibility and magnetization measurements. We have shown that in both
samples a minority portion of the clusters relaxes faster than the other ones
and that its different relaxation arises from a broad distribution of energy
barriers lower than that of the majority portion. With the use of ac magnetic
measurements, we have estimated the relative amount of these minority species
together with the temperature and field dependences of their effective
relaxation times. In the temperature range of 2~K, we have shown that the
relaxation of the secondary species is the dominant one in relaxation
experiments, at least at short times. In the case of Mn$_{12}$~2-Cl benzoate,
we have also shown that the minority species experiences spin tunneling
between excited states at zero applied field.

\ack We are indebted with J. Tejada and J. M. Hern\'andez for supplying the
samples used in this work and with J. F.  Fern\'andez and D. Ruiz for useful
discussions. M. E. acknowledges the European Community for a TMR Grant under
network contract No. ERBFMRXCT980181. This work was partially supported by
Grant No. MAT96/448 from CICYT.

\pagebreak

\begin{figure}[!h]
\caption{Plot of the zero-field out-of-phase (imaginary) susceptibility versus
temperature collected at $f=620$~Hz for Mn$_{12}$ acetate $(\bullet)$ and for
Mn$_{12}$~2-Cl benzoate $(\circ)$. In the inset, the magnification for
$1.5~{\rm{K}}<T<4~\mathrm{K}$. The lines are visual guides.}
\end{figure}

\begin{figure}[!h] 
\caption{Zero-field differential susceptibilities in the complex plane at
$T=1.70$~K for Mn$_{12}$ acetate $(\bullet)$ and for Mn$_{12}$~2-Cl benzoate
$(\circ)$. The data are normalized with their respective isothermal values.
The lines are the fits to semi-circles.}
\end{figure}

\begin{figure}[!h]
\caption{Semi-logarithmic plot of the zero-field relaxation times of
Mn$_{12}$ acetate (squares) and Mn$_{12}$~2-Cl benzoate (circles) with data
taken from Ref.~\cite{mncl2} (filled symbols), and from the present work
(open symbols). The straight lines are the Arrhenius fits.}
\end{figure}

\begin{figure}[!h] 
\caption{Experimental dependences of the minority species relaxation times on
the applied field at $T=1.7$~K for the Mn$_{12}$ acetate and at $T=2.0$~K for
the Mn$_{12}$~2-Cl benzoate. The continuous lines are the estimated
dependences for classical over-barrier hopping calculated for $T=1.7$~K,
$\tau_{0}=2\times 10^{-9}$~s, $D=0.35$~K, $S=10$ and $T=2.0$~K,
$\tau_{0}=10^{-9}$~s, $D=0.25$~K, $S=10$, for the acetate and the benzoate,
respectively.}
\end{figure}

\begin{figure}[!h]
\caption{Time decay of the magnetization, measured at zero applied field after
saturation for the indicated temperatures, plotted as a function of the
logarithm of the time for the Mn$_{12}$ acetate ($a$) and the Mn$_{12}$~2-Cl
benzoate ($b$). Straight lines are linear fits. In the insets: same data but
plotted as a function of the square root of the time.}
\end{figure}


\begin{thebibliography}{99}
\bibitem{testo} Early investigations~\cite{paulsen} on the magnetic
relaxation at $T<2$~K show that the relaxation time saturates at values of
$10^{7}$-$10^{8}$~s. In contrast, in Ref.~\cite{thom}, it is shown that, below
2~K, the zero-field relaxation  of the magnetic moment has a square root decay
and that the extrapolated relaxation time depends weakly on the temperature.
Other recent magnetization  measurements show that the relaxation is
logarithmic below 1~K~\cite{pere}.
\end{thebibliography}
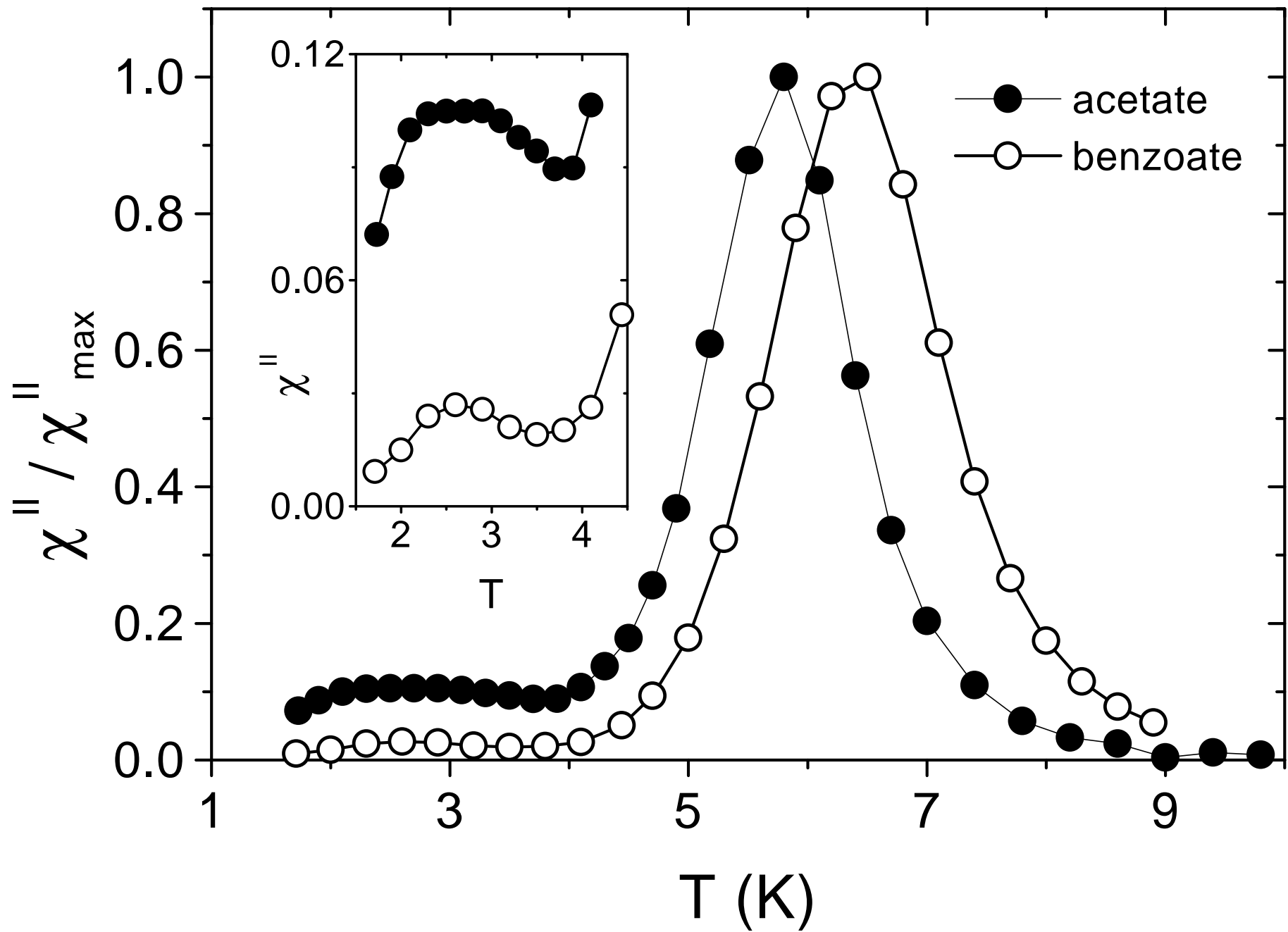
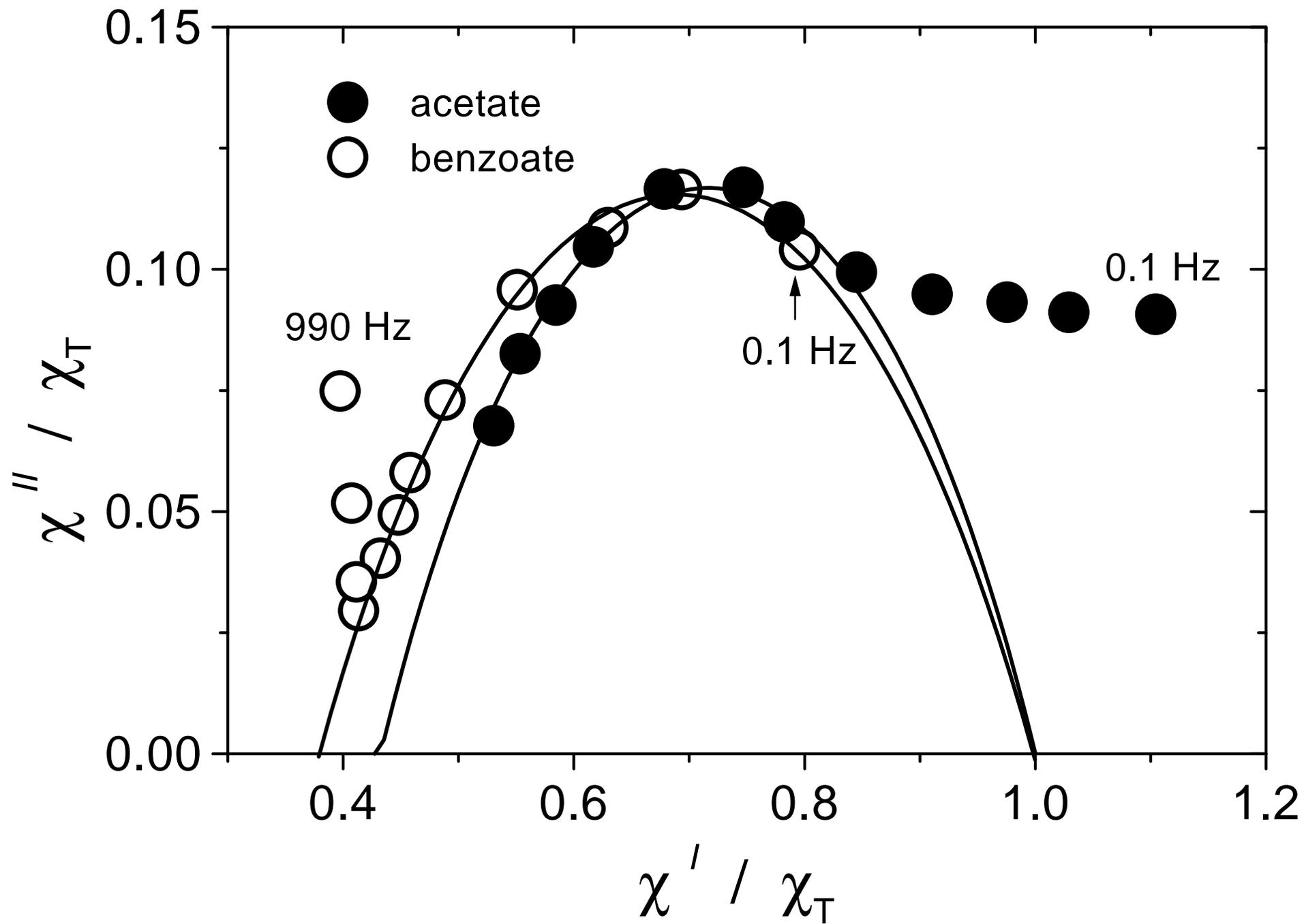
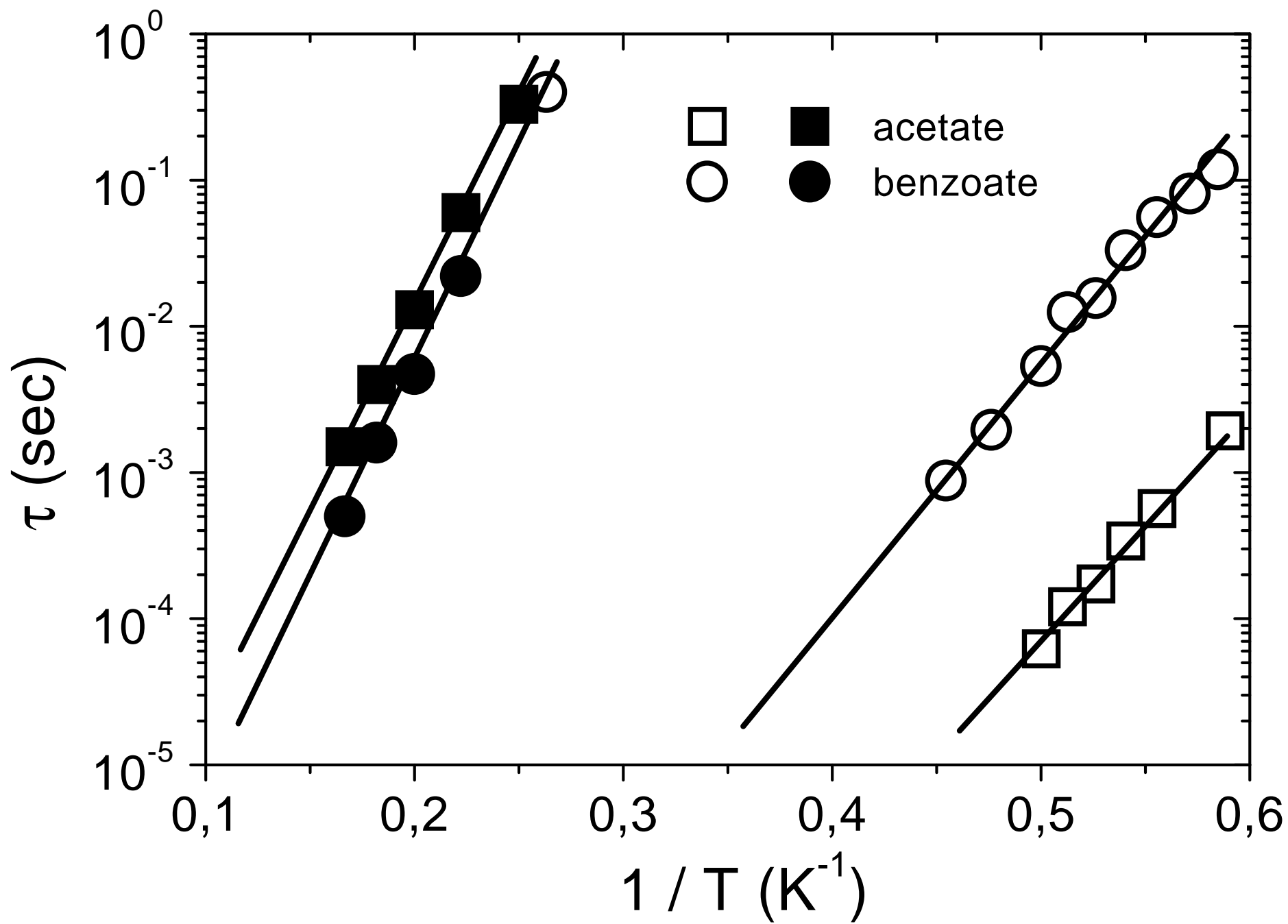
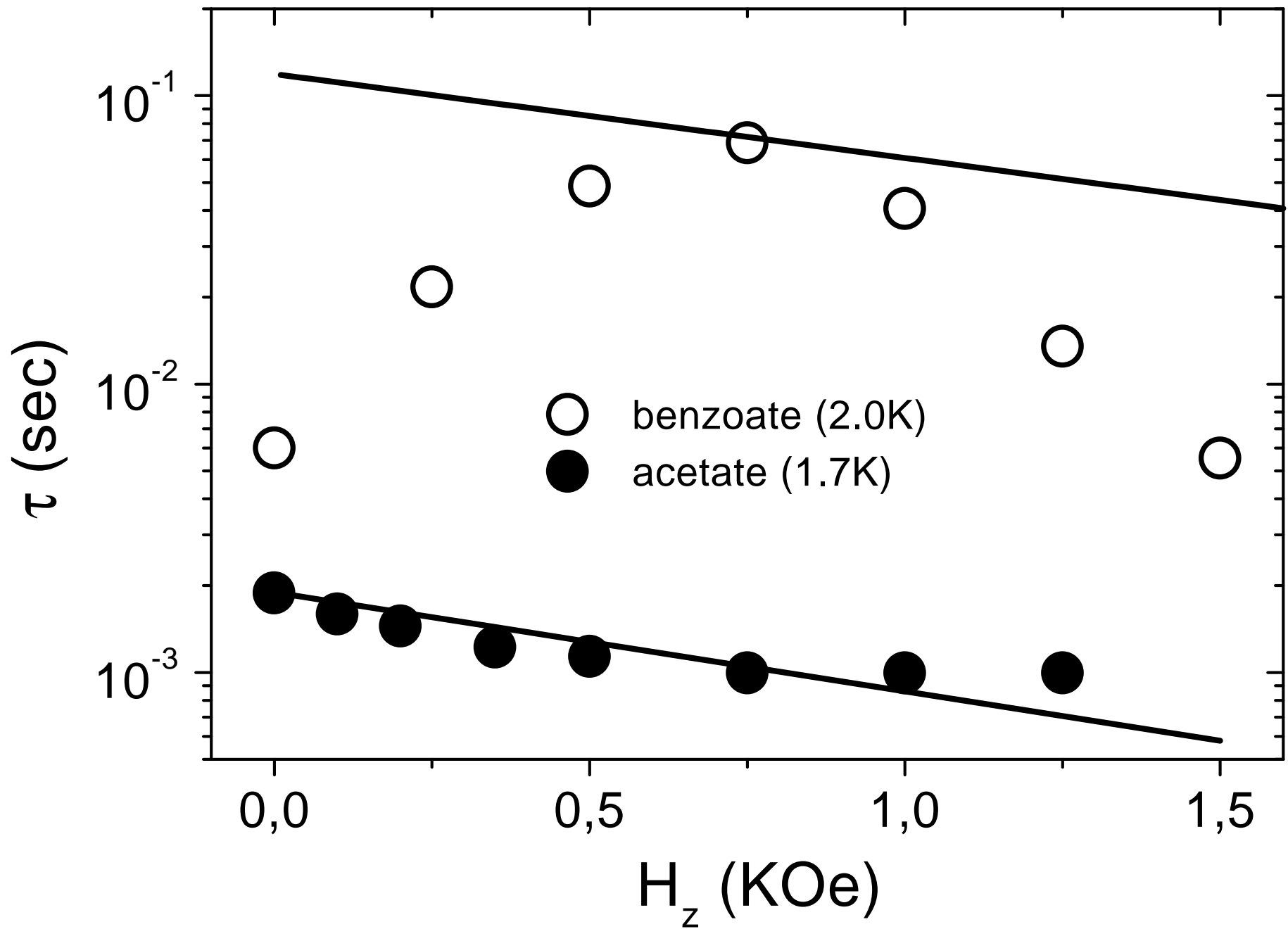
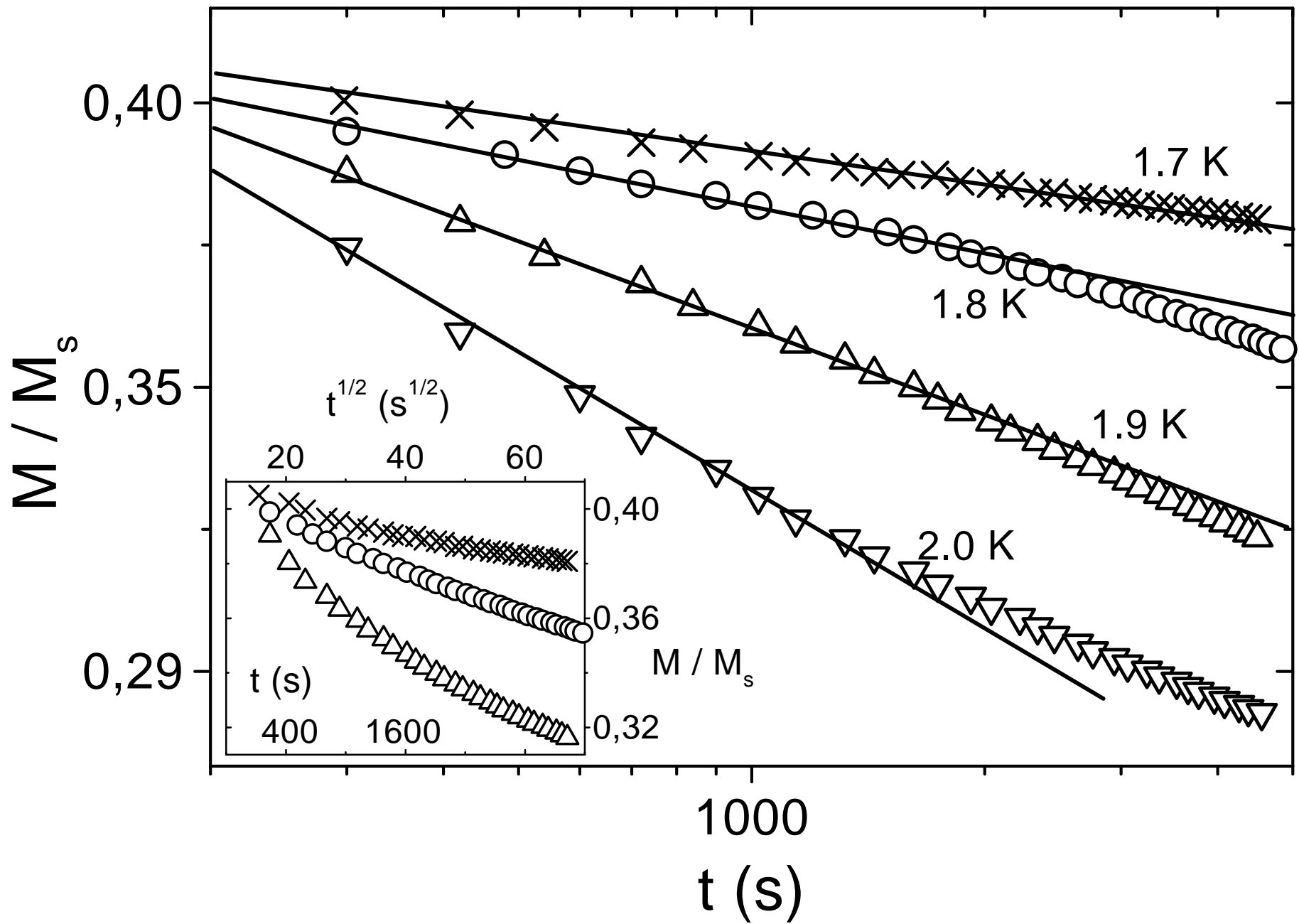
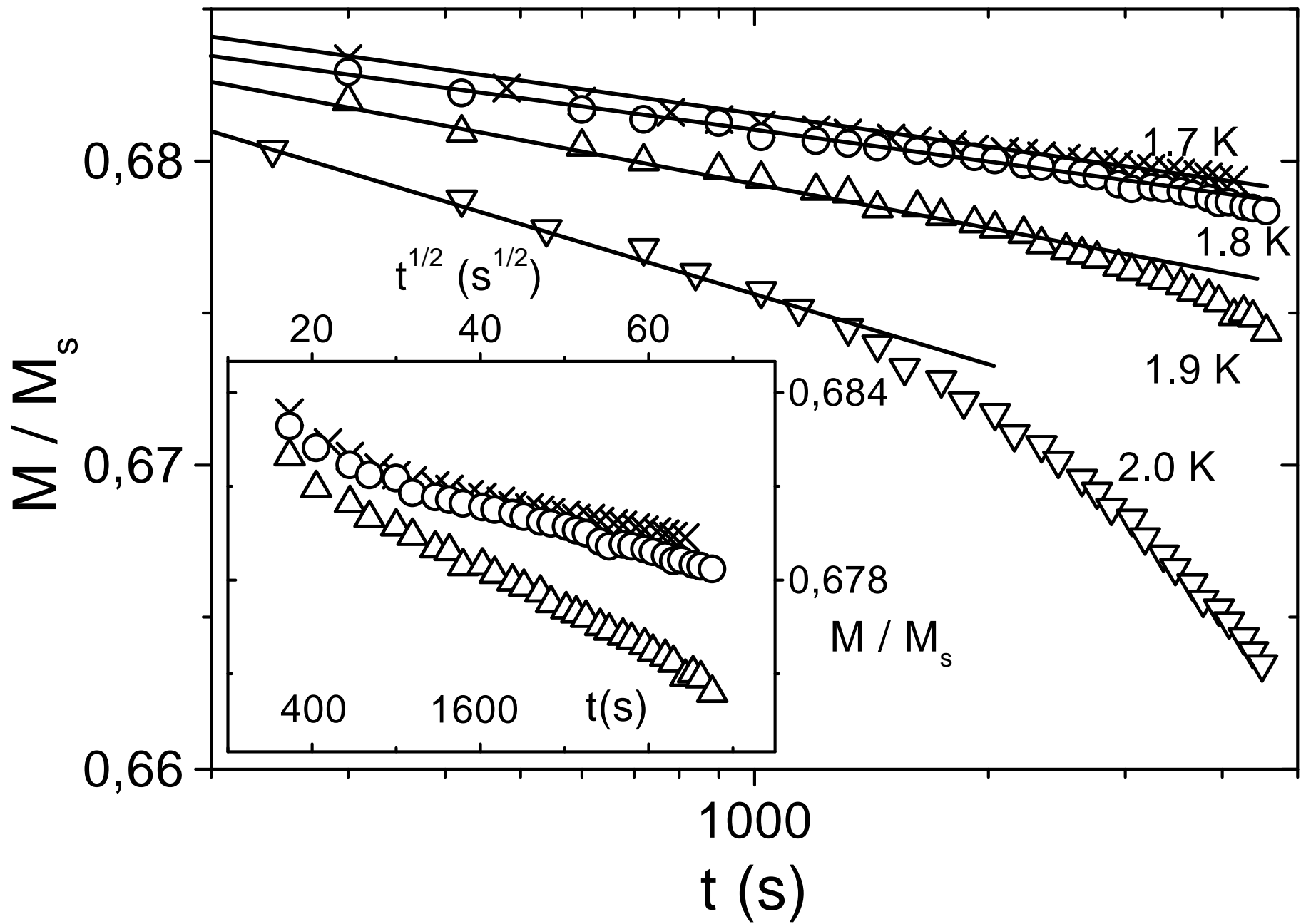
\end{document}